\documentclass[structabstract]{aa}
\usepackage{graphicx} 
\usepackage{pdflscape} 
\usepackage{txfonts}
\usepackage{natbib}
\bibpunct{(}{)}{;}{a}{}{,} 
\usepackage{amssymb}
\begin{document}
%
\newcommand{\figref}[1]{Fig.~\ref{#1}} 
\newcommand{\up}[1]{$^{#1}$}
\newcommand{\down}[1]{$_{#1}$}
\newcommand{\tblmark}[1]{\tablefootmark{#1}}
\newcommand{\tabref}[1]{Table~\ref{#1}}
\newcommand{\spitzer}{\textit{Spitzer}}
\newcommand{\herschel}{\textit{Herschel}}
\newcommand{\tento}[1]{$\times10^{#1}$}
\newcommand{\textsim}{$\sim$}
\newcommand{\kms}{km~s$^{-1}$}
\newcommand{\mjykms}{mJy~\kms}
\newcommand{\jykms}{Jy~km~s$^{\rm-1}$}
\newcommand{\Msun}{$M_{\rm\sun}$}
\newcommand{\Lsun}{$L_{\rm\sun}$}
\newcommand{\Mearth}{$M_{\rm\oplus}$}
\newcommand{\water}{{H$_2^{18}$O}}
\newcommand{\normalwater}{H$_2$O}
\newcommand{\dimethylether}{CH$_3$OCH$_3$}
\newcommand{\ethylcyanide}{C$_2$H$_5$CN}
\newcommand{\sulfurdioxide}{SO$_2$}
\newcommand{\trans}[6]{$#1_{#2,#3}-#4_{#5,#6}$}
\newcommand{\htwootrans}{$3_{1,3}-2_{2,0}$}
%
   \title{Subarcsecond resolution observations of warm water towards
     three deeply embedded low-mass protostars\thanks{Based on
   observations carried out with the IRAM Plateau de Bure
   Interferometer. IRAM is supported by INSU/CNBRS(France), MPG
   (Germany) and IGN (Spain).}}

   \author{M.~V. Persson
          \inst{1}\fnmsep\inst{2}
          \and
          J.~K. J{\o}rgensen\inst{2}\fnmsep\inst{1}
          \and E.~F. van Dishoeck
          \inst{3}\fnmsep\inst{4}
          }

   \institute{Centre for Star and Planet Formation,
   Natural History Museum of Denmark,
   University of Copenhagen,
   {\O}ster Voldgade 5-7, \\DK-1350,
  Copenhagen K, Denmark\\
              \email{magnusp@nbi.dk}
         \and
            Niels Bohr Institute, University of Copenhagen, Juliane
            Maries Vej 30, DK-2100 Copenhagen {\O}, Denmark
         \and
             Leiden Observatory, Leiden University, P.O. Box 9513, NL-2300 RA Leiden, The Netherlands
        \and
            Max-Planck Institute f\"{u}r extraterrestrische Physik (MPE), Giessenbachstrasse, 85748 Garching, Germany
             }

   \date{Received ; accepted }


  \abstract
{Water is present during all stages of star formation: as ice in the
  cold outer parts of protostellar envelopes and dense inner regions
  of circumstellar disks, and as gas in the envelopes close to the
  protostars, in the upper layers of circumstellar disks and in
  regions of powerful outflows and shocks. Because of its key
  importance in the understanding of its origin in our own
  solarsystem, following the evolution of water all the way to the
  planet-forming disk is a fundamental task in research in star
  formation and astrochemistry.}
  {In this paper we probe the mechanism regulating the warm gas-phase
    water abundance in the innermost hundred AU of deeply embedded
    (Class~0) low-mass protostars, and investigate its chemical
    relationship to other molecular species during these stages.}
  {Millimeter wavelength thermal emission from the para-\water\
    \htwootrans\ ($E_\mathrm{u}$=203.7~K) line is imaged at high angular
    resolution (0\farcs75; 190~AU) with the IRAM Plateau de Bure
    Interferometer towards the deeply embedded low-mass protostars
    NGC~1333-IRAS2A and NGC~1333-IRAS4A.}
{Compact \water\ emission is detected towards IRAS2A and one of the
  components in the IRAS4A binary; in addition \dimethylether,
  \ethylcyanide, and \sulfurdioxide\ are detected. Extended water
  emission is seen towards IRAS2A, possibly associated with the outflow.}
{The results complement a previous detection of the same transition
  towards NGC~1333-IRAS4B. The detections in all systems suggests that
  the presence of water on $\lesssim$~100~AU scales is a common
  phenomenon in embedded protostars and that the non-detections of hot
  water with \spitzer\ towards the two systems studied in this paper
  are likely due to geometry and high extinction at mid-infrared
  wavelengths. We present a scenario in which the origin of the
  emission from warm water is in a flattened disk-like
  structure dominated by inward motions rather than rotation. The gas-phase water abundance
  varies between the sources, but is generally much lower than a
  canonical abundance of 10\up{-4}, suggesting that most
  water (\textgreater96\%) is frozen out on dust grains at these
  scales. The derived abundances of \dimethylether\ and
  \sulfurdioxide\ relative to \water\ are comparable for all sources
  pointing towards similar chemical processes at work. In contrast, the
  \ethylcyanide\ abundance relative to \water\ is significantly lower
  in IRAS2A, which could be due to different chemistry in the
  sources.} 
    

   \maketitle


\section{Introduction} Water is a key molecule in regions of star 
formation. In cold outer regions of protostellar envelopes and dense
mid-planes of circumstellar disks, it is the most abundant molecule in
the ices on dust grains \citep[e.g.][]{gibb2004,boogert2008}.
In regions of increased temperatures (e.g., due to protostellar
heating or active shocks) it desorbs and becomes an important coolant
for gas \citep{ceccarelli1996, neufeld1993}. As one of the major
oxygen reservoirs it also influences the oxygen based chemistry
\citep[e.g.][]{dishoeck1998}. This paper presents high-angular
resolution observations (0\farcs75) at 203.4~GHz of the
\water\ isotopologue towards two deeply embedded (Class~0) low-mass
protostars using the Institute de Radioastronomie Millim\'{e}trique
(IRAM) Plateau de Bure Interferometer (PdBI), one of the few water
lines that can be routinely observed from Earth. The aim of the
observations is to determine the origin of water emission lines in the
inner 100~AU of these protostars.

The young, dust- and gas-enshrouded protostars provide an important
laboratory for studies of the star formation process. On the one hand,
they provide the link between the prestellar cores and the young
stellar objects (YSOs) and on the other hand it is likely that during
these stages the initial conditions for the later astrochemical
evolution of the protostellar envelopes and disks are established
\citep[e.g.,][]{visser2009}. In the earliest stages most of the gas
and dust is present on large scales in the protostellar envelopes
where the temperatures are low (10--20~K) and most molecules,
including water, are locked up as ice on grains. Nevertheless,
space-based observations from \textit{ISO, SWAS, Odin,} \spitzer\ and
\herschel\ show that significant amounts of gas-phase water are
present towards even deeply embedded protostars, pointing to
significant heating and processing of water-ice covered grains
\citep{nisini1999,ceccarelli1999,maret2002,bergin2003,franklin2008,watson2007,dishoeck2010,kristensen2010}.

Exactly which mechanism regulates where and how the heating and
evaporation of the water ice takes place is still debated. Proposed
locations and mechanisms include the innermost envelope, heated to
temperatures larger than 100~K by the radiation from the central
protostar on scales $\lesssim$100~AU \citep{ceccarelli1998,maret2002},
in shocks of the protostellar outflow \citep{nisini1999}, or in the
disk -- heated either by shocks related to ongoing accretion onto the
circumstellar disks or in their surface layers heated by the protostar
\citep{watson2007,jorgensen2010a}.

Recently, \citet{watson2007} observed 30 Class 0 protostars, including
the sources in this paper, with the infrared spectrograph (IRS) on
board \spitzer. The protostar NGC~1333-IRAS4B (IRAS4B) showed a wealth
of highly excited \normalwater\ lines, and was the only source in the
sample in which the water lines were detected. By modeling the spectra
they concluded that because of the high inferred density and
temperature of the water, it originates in the vicinity of the
protostar, likely accretion shocks in the circumstellar disk.
However, the \spitzer\ data, and similar observations from space, lack
the spatial and spectral resolution to unambiguously associate the
emission with the components in the protostellar systems -- and can
for the shorter wavelengths also be hampered by extinction by the dust
in the protostellar envelopes and disks. \citet{herczeg2011} analyzed
the emission from warm water lines in maps toward IRAS4B obtained with
the PACS instrument on the \herschel\ \textit{Space Observatory}. The maps show
that the bulk of the warm water emission has its origin in a
outflow-driven shock offset by about 5\arcsec\ (1200~AU) from the
central protostar. An excitation analysis of the emission furthermore
shows that the temperature in the shocked region is about 1500~K,
which can simultaneously explain both the far-IR (\herschel) and
mid-IR (\spitzer) water lines.

Higher angular and spectral resolution is possible through
interferometric observations of \water\ which has a few transitions
that can be imaged from the ground
\citep[e.g.][]{jacq1988,gensheimer1996,tak2006}. \citet{jorgensen2010a}
report observations of IRAS4B with the PdBI. They detect the
\water\ \htwootrans\ line at 203.4~GHz, and also resolve a tentative
velocity gradient in the very narrow line {($\sim$2~\kms\ width at
  zero intensity),} indicating an origin of the \water\ emission in a
rotational structure. The \normalwater\ column density is around 25
times higher than previously estimated from mid-IR observations by
\citet {watson2007}, on similar size scales (inner 25~AU of the
protostar).

This study presents similar high angular resolution H$_2^{18}$O
millimeter wavelength observations for the sources NGC~1333-IRAS2A
(IRAS2A) and NGC~1333-IRAS4A (IRAS4A), which are compared to
IRAS4B. The three sources are all deeply embedded Class~0 protostars
in the nearby NGC~1333 region of the Perseus molecular
cloud\footnote{In this paper we adopt a distance of 250~pc to the
  Perseus molecular cloud \citep[][and references
    therein]{enoch2006}.} identified through early infrared and
(sub)millimeter wavelength observations
\citep[e.g.,][]{jennings1987,sandell1991}. All sources were mapped by
\spitzer\ \citep {jorgensen2006, gutermuth2008}, and were part of the
high resolution ($\sim$1\arcsec) PROSAC survey of Class 0/I sources
using the Submillimeter Array (SMA; \citealt {jorgensen2007,
  jorgensen2009}). Their similar luminosities (20,~5.8\footnote{The
  luminosity and mass estimates for IRAS4A refer to the binary
  system.},~3.8~\Lsun\ for IRAS2A, IRAS4A and IRAS4B, respectively)
and envelope masses (1.0,~4.5$^2$,~2.9~\Msun, respectively) from those
surveys, and thus presumably also similar evolutionary stages, make
them ideal for studying and comparing deeply embedded young stellar
objects in the same environment.

IRAS4A is a binary system with a 1\farcs8 separation \citep
{looney2000,reipurth2002, girart2006}, associated with a larger scale
jet and outflow \citep[e.g.][]{blake1995, jorgensen2007}. The jet axis
has been deduced from water maser observations associated to IRAS4A-NW
to $2\degr$ in the plane of the sky and at a position angle (PA) of
$-50\degr$ \citep[north through east,][]{marvel2008,desmurs2006},
almost perpendicular to the CO outflow axis (PA$=0\degr-45\degr$;
\citealt{blake1995,jorgensen2007}). The mass of the NW source is
estimated from modeling to 0.08~\Msun\ \citep{choi2010}, no estimate
of the mass of the SE source exists. The envelope is dominated by
inward motion, as evident from the inverse P-Cygni profiles in the
spectra and modeling of gas kinematics
\citep{francesco2001,attard2009,kristensen2010}.

The other source, IRAS2A, shows two bipolar outflows, one shell-like
directed N-S \citep[PA$\approx+25\degr$,][]{jorgensen2007} and the
other a jet-like directed E-W
\citep[PA$\approx+105\degr$,][]{liseau1988,sandell1994,jorgensen2004}. Water
maser emission has been detected towards IRAS2A, partly associated with
the northern redshifted outflow \citep{furuya2003}. One proposed
solution to the presence of a quadrupolar outflow is that IRAS2A is a
binary system with 0\farcs3 (75~AU) separation \citep{jorgensen2004}.

The paper is laid out as follows. Sect.~\ref{sec:obs} presents the
interferometric observations of the three Class 0 sources that the
paper is based upon. Sect.~\ref {sec:res} presents the results from
the analysis of the spectra and maps. Sect.~\ref{sec:dis} discusses
the detection of water, and the column density
estimates. Sect.~\ref{sec:sum} presents a summary of the important
conclusions and suggests possible future work.


\section{Observations}\label{sec:obs} Three low-mass protostars,
IRAS2A, IRAS4A-SE and -NW in the embedded cluster NGC~1333 in the
Perseus molecular cloud were observed using the PdBI. IRAS2A and
IRAS4A were observed in two configurations each: the B configuration
on 6, 15 and 17 March 2010 and the C configuration on 9 December
2009. The sources were observed in track-sharing mode for a 
total of 12 hours in B configuration and about 8 hours in C configuration. 
Combining the observations in the different configurations, the data cover baselines
from 15.4 to 452~m (10.5 to 307~k$\lambda$). The receivers were tuned
to the \mbox {para-\water} \htwootrans\ transition at 203.40752~GHz
(1.47~mm) and the correlators were set up with one unit with a
bandwidth of 40~MHz (53~\kms) centered on this frequency providing a
spectral resolution on 460 channels of 0.087~MHz (0.115~\kms) width.

The data were calibrated and imaged using the CLIC and MAPPING
packages, which are parts of the IRAM GILDAS software. The general
calibration procedure was followed: regular observations of the nearby,
strong quasar 0333+3221 were used to calibrate the complex gains and
bandpass, while MWC349, 0333+3221 and 3C84 were observed to calibrate
the absolute flux scale. Integrations with significantly deviating
amplitudes/phases were flagged and the continuum was subtracted before
Fourier transformation of the data. The resulting beam sizes using
natural weighting, along with other data, are given in
\tabref{table:obs}. The field of view is 25\arcsec (HPBW) at 1.45~mm
and the continuum sensitivity is limited by the dynamical range of the
interferometer while the line data RMS is given in
\tabref{table:obs}. The uncertainty in fluxes is dominated by
  the flux calibration error, typically 20\%.

\begin{table}[h]

    \caption{Observation parameters.}

    \label{table:obs}      
    \centering                          
        \begin{tabular}{llll}        
            \hline\hline                 
                Source          & Beam\tblmark{a} & PA & RMS\tblmark{b}  \\  
                \hline                        
                IRAS2A          & 0\farcs87~x~0\farcs72 & 63.5\degr & 16.1\\ 
                IRAS4A(-SE/NW)  & 0\farcs86~x~0\farcs70 & 63.4\degr & 16.5  \\
                \hline
        \end{tabular}

        \tablefoottext{a}{Major $\times$ minor axis.}
        \tablefoottext{b}{Units of mJy~beam$^{-1}$~channel$^{-1}$.}
\end{table}


\section{Results}\label{sec:res}

Continuum emission from the targeted sources (IRAS2A and the IRAS4A-NW
and IRAS4A-SE binary) was detected at levels expected from previous
high resolution millimeter wavelength continuum observations:
parameters from elliptical Gaussian fits to the continuum emission for
the sources are given in \tabref{table:cont_fits}. The continuum
emission towards the sources is resolved and the fluxes concur with
the results of \citet{jorgensen2007}, assuming a power-law spectrum
($F_{\nu}$~$\propto$~$\nu^{\alpha}$) with $\alpha$$\sim$2.3--2.7, expected 
for thermal dust continuum emission.

\begin{table*}
\centering
\begin{minipage}{1.4\columnwidth}
   \caption{Parameters for the sources from elliptical Gaussian fits
      to their continuum emission. Errors given are the statistical
      errors from the fitting routine.}

    \label{table:cont_fits}
       \begin{tabular}{llll}
            \hline\hline
                                       & IRAS2A                                 & IRAS4A-NW                                     & IRAS4A-SE \\
                \hline
                Flux ($\pm$)\tblmark{a}& $0.24$ (0.002) Jy                      & $0.88$ (0.025) Jy                                  & $1.46$ (0.021) Jy\\
                R.A ($\pm$)\tblmark{b} & 03:28:55.57 (0\farcs004)               & 03:29:10.44 (0\farcs02)                             & 03:29:10.53 (0\farcs005)  \\
                Dec ($\pm$)\tblmark{b} & 31:14:37.03 (0\farcs004)               & 31:13:32.16 (0\farcs02)                             & 31:13:30.92 (0\farcs006) \\
                Extent\tblmark{c}      & $1\farcs1\times1\farcs0~(23^{\circ})$  & $2\farcs0\times1\farcs2~(-53^{\circ})$    & $1\farcs4\times1\farcs0~(50^{\circ})$ \\
            \hline
          \end{tabular}\\ 
          \tablefoottext{a}{ Statistical error from fit, the
            calibration dominates the uncertainty in the flux
            determination.}  
          \tablefoottext{b}{J2000 epoch.  Statistical error from fit.}
          \tablefoottext{c}{Major $\times$ Minor axis and position
            angle for elliptical Gaussian profile in $(u,v)$ plane. The
            statistical error of the extent is typically $0.5-3\%$.}
\end{minipage} 
\end{table*}

\begin{figure}
  \centering
        \includegraphics[width=\columnwidth]{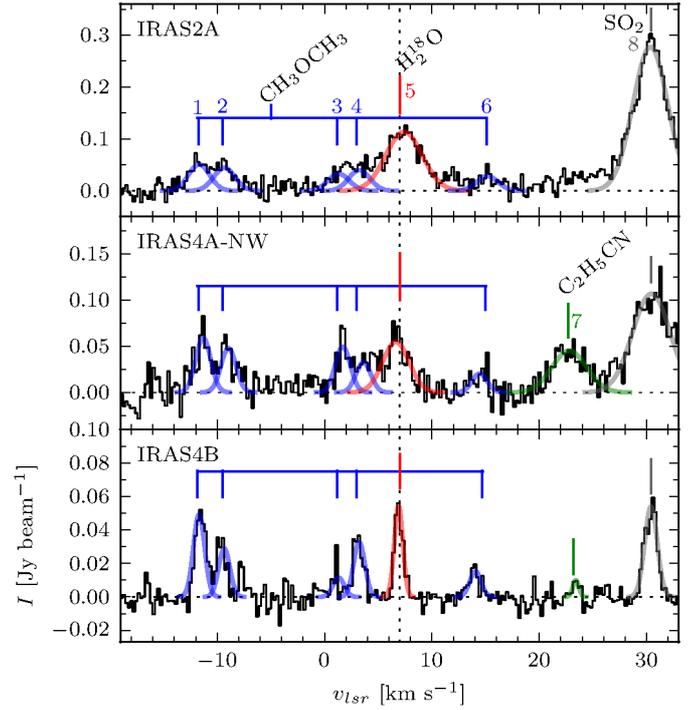}
        \caption{\label{fig:spectra} Spectra extracted in a pixel
            towards the continuum peak positions for IRAS2A, IRAS4A-NW
            and IRAS4B. The identified lines are marked, including the
            targeted \water\ \htwootrans\ line. Parameters from the fits are tabulated in 
	Table~\ref{tab:linefits}.2. The numbers above the
            lines and the colors correspond to the numbers (emission
            lines) and colors in \figref{fig:linepos},
            Table~\ref{tab:intensities},
            Table~\ref{tab:molecules} and Table~\ref{tab:linefits}.2. 
	The spectra have been binned to
            a width of 0.23~\kms\ (i.e., 2 channels per bin), and
            Gaussian profiles have been fitted to the identified
            lines. The systemic velocity of the \water\ line is marked
            by the vertical dotted line at $v_\mathrm{
              lsr}=7.0$~\kms. Note the difference in scale of the {\it
              y}-axis for the spectra.}
\end{figure}

\figref{fig:spectra} shows the spectrum towards the continuum peak of
these sources, as well as IRAS4B \citep{jorgensen2010a}.  A number of
lines are detected towards IRAS2A and one component in the IRAS4A
protobinary system, IRAS4A-NW. IRAS4A-SE in contrast does not show any
emission lines in the high-resolution spectra, and it is therefore
only listed in the table showing the characteristics of the continuum
emission.

The Jet Propulsion Laboratory \citep[JPL,][]{pickett1998} and the
Cologne Database for Molecular Spectroscopy \citep
[CDMS,][]{muller2001} databases were queried through the Splatalogue
interface to identify the lines. The line widths and strengths vary
between the sources and because of the broader line widths some
blending between the different \dimethylether\ lines occurs in both
IRAS2A and IRAS4A-NW. As seen in \figref{fig:spectra}, several lines
are detected: the target water line, {\sulfurdioxide} and several
lines of {\dimethylether}. \ethylcyanide\ is seen in IRAS 4A-NW and
IRAS4B, but not in IRAS2A.

Particular interesting for this study is the assignment of the
\water\ \htwootrans\ transition ($E_\mathrm{u}$=203.7~K). No other
lines fall within $\pm$1~\kms\ of the \water\ \htwootrans\ line listed
in the spectral line catalogs. The lines that fall close
($\pm$1-2~\kms) are {\up{13}CH\down{2}CHCN},
{\up{13}CH\down{3}CH\down{2}CN} and {(CH\down{3})\down{2}CO}. Both the
{\up{13}CH\down{2}CHCN} and {\up{13}CH\down{3}CH\down{2}CN}
isotopologues have other transitions which would have been detected,
whereas a possible transition of {(CH\down{3})\down{2}CO} has a too
low intrinsic line strength to be plausible.

\begin{figure*}[ht]
\sidecaption \includegraphics[width=120mm]{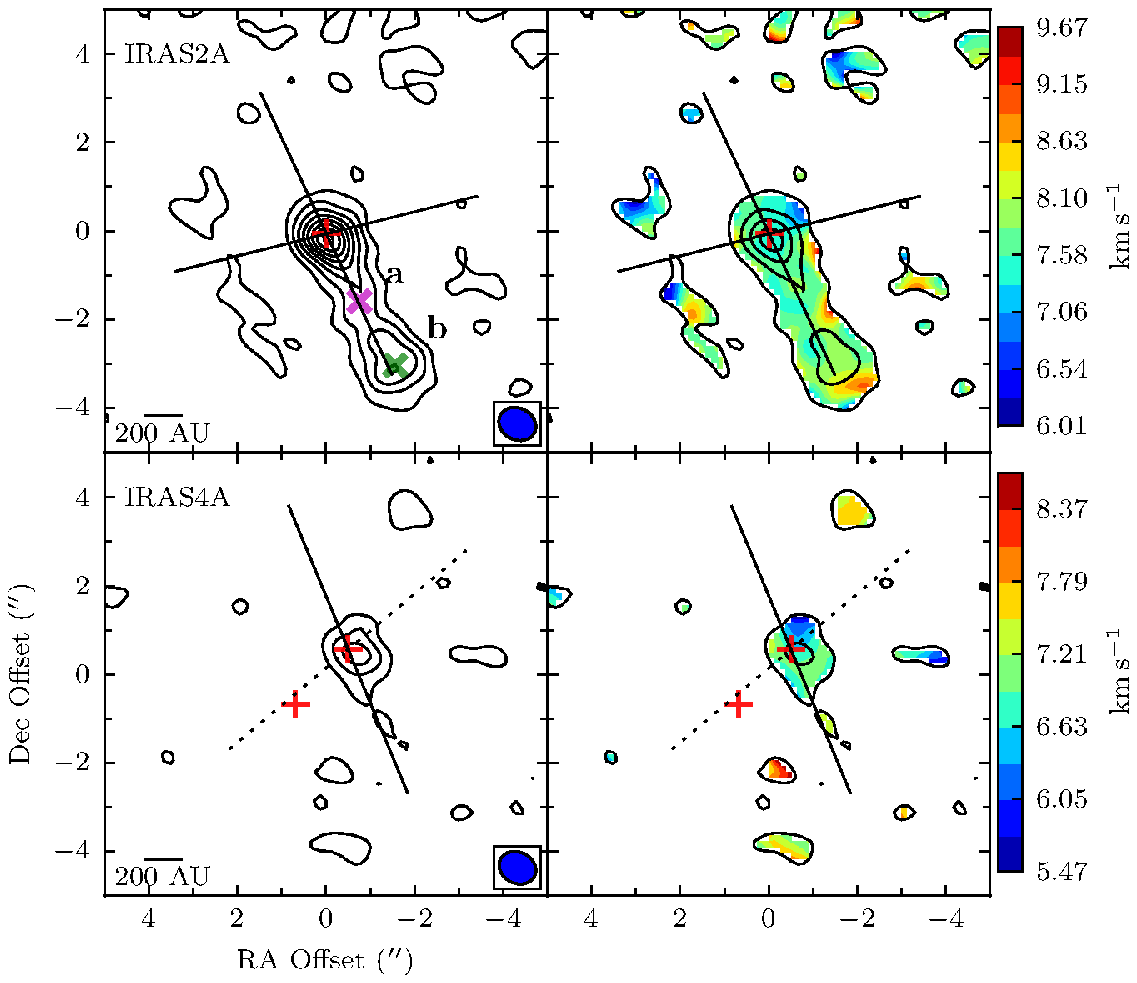}
    \caption{\label{fig:mom0} Integrated intensity (left) and velocity
      (right) maps for the \water~\htwootrans\ line in IRAS2A
      (from 4.4 to 11.4~\kms) and IRAS4A (from 4.7 to
        9.6~\kms). Contours show integrated intensity and are in
      steps of 4$\sigma$ for the intensity map, 8$\sigma$ for the
      velocity map, and start at 3$\sigma$. The position of the
      continuum emission is marked by a red plus sign, an outflow by
      the solid line and maser emission by the dashed line. Positions
      where spectra have been extracted in IRAS2A along the outflow
      are marked with letters {\bf a} and {\bf b} and crosses with
      magenta and green colors in the integrated intensity map.}
\end{figure*}

In \figref{fig:mom0} the integrated intensity (zeroth order moment)
and velocity (first order moment) maps for the \water\ transition are
shown for the observed sources. For IRAS2A the calculations were
  made over the velocity interval from 4.4 to 11.4~\kms\ and for
  IRAS4A-NW from 4.7 to 9.6~\kms. The integrated intensity maps for
  IRAS2A show some extended emission, whereas for IRAS4A-NW
  the emission is compact around the continuum peak with a slight
  elongation in the north-south direction. The velocity maps for both
IRAS2A and IRAS4A (right hand panels in \figref{fig:mom0}) show no
obvious spatial gradients in velocity, neither on source nor along the
outflow in IRAS2A.

For estimates of the physical conditions in the gas, circular
  Gaussian profiles were fitted to the emission of each line in the
  $(u,v)$-plane. The fits were performed to the line intensities typically
  integrated over the line widths ($v_{\rm lsr}\pm$FWHM) and are given
  in Table~\ref{tab:intensities}. Table~\ref{tab:molecules} lists the
  molecular parameters adopted in this paper of the target line
  \water\ \htwootrans, and other detected lines, while Table~\ref{tab:linefits}.2
lists the position, FWHM and integration interval for the different lines.

\begin{table*}

\begin{minipage}{0.79\linewidth}
    \caption{\label{tab:intensities} Parameters for the lines
        detected in the sources from circular Gaussian fits to the
        integrated line maps. Molecule specific data are tabulated in
        Table~\ref{tab:molecules} and the integration interval (in \kms) 
 	in Table~\ref{tab:linefits}.2 in the appendix. }

\begin{tabular}{lllccccccccc}
\hline\hline
         &                                    &                                                           	&                                         & IRAS2A                	&                                    		& &                    			& IRAS4A-NW         	&  \\
\cline{4-6}\cline{8-10}                                                                                                             
No.      & Molecule            & Transition                                       	& Flux\tblmark{a}        & Offset\tblmark{b} & Size\tblmark{c}        		& & Flux\tblmark{a} 		& Offset\tblmark{b} 	& Size\tblmark{c}  \\
           &                                 &                                                             	& (Jy\,\kms)                   & (\arcsec)                    & (\arcsec)                     		& & (Jy\,\kms)      		& (\arcsec)         		& (\arcsec)  \\
\hline                                                                                                                  
1        &\dimethylether  & 3\down{3,0} -- 2\down{2,1} EE    	& $0.47$\tblmark{d}   & $-0.02;-0.002$     	& $0.70\pm0.09$    		& & $0.44$\tblmark{d}  	& $-0.21;-0.04$     	& $0.70\pm0.09$ \\
2        &\dimethylether  & 3\down{3,0} -- 2\down{2,1} AA   	& $-$\tblmark{d}         	& $-;-$                       	& $-$                           		& & $-$\tblmark{d}      	& $-;-$                           	& $-$    \\
3        &\dimethylether  & 3\down{3,0} -- 2\down{2,1} AE    	& $0.81$\tblmark{e}  	& $-0.16;-0.32$       	& $1.45\pm0.10$     		& & $0.26$\tblmark{e} 	& $-0.14;-0.03$    	& $0.57\pm0.12$ \\
4        &\dimethylether  & 3\down{3,1} -- 2\down{2,1} EE     	& $-$\tblmark{e}        	& $-;-$                        	& $-$                           		& & $-$\tblmark{e}       	& $-;-$                            	& $-$    \\
5        &\water                    & 3\down{1,3} -- 2\down{2,0}           	& $0.98$                         	& $-0.07;-0.12$       	& $0.83\pm0.05$    		& & $0.27$                 		& $-0.18;-0.07$           	& $0.61\pm0.12$ \\
6        &\dimethylether  & 3\down{3,1} -- 2\down{2,1} EA    	& $0.30$                         	& $-0.35;-0.52$       	& $1.23\pm0.22$     		& & $0.06$       	            	& $-0.25;-0.10$           	& $0.44\pm0.26$ \\
7        &\ethylcyanide     & 23\down{2,22} -- 22\down{2,21} 	& $<0.04$\tblmark{f} 	& $-;-$                        	& $-$                          		& & $0.27$                		& $-0.20;-0.05$     	& $0.37\pm0.11$ \\
8        &\sulfurdioxide     & 12\down{0,12} -- 11\down{1,11}	& $2.73$                          & $-0.07;+0.003$    	& $1.00\pm0.02$     		& & $1.22$                      	& $-0.22;-0.01$            	& $0.88\pm0.05$ \\
\hline
\end{tabular}\\

\tablefoottext{a}{Flux from fitting circular Gaussian to integrated
  line emission in ($u,v$) plane. The typical statistical error is
  0.01~Jy~\kms, but the calibration uncertainty is roughly 20\%. }

\tablefoottext{b}{Offset from position of continuum fitted elliptical
  Gaussian. Used in \figref{fig:linepos} to plot the positions of the
  lines.}

\tablefoottext{c}{FWHM from Gaussian fit in the ($u,v$) plane.}

\tablefoottext{d,e}{Blended pair, values given for the combined
  integrated emission.}

\tablefoottext{f}{Upper limit estimate.}

\end{minipage}
\end{table*}


\section{Analysis and discussion}\label{sec:dis} The detected
  water lines provide important constraints on the physical and
  chemical structure of the deeply embedded low-mass protostars, which
  we discuss in this section. The first part of the discussion
  (Sect.~\ref{sec:smallScales}--\ref{sec:geometry}) focuses on the
  detection of warm water on scales comparable to the protostellar
  disk and the extended emission found in the outflow towards IRAS2A.
  The implications for the geometry and dynamics are also
  discussed. The second part
  (Sect.~\ref{sec:masing}-\ref{sec:formation}) discusses excitation
  conditions in the water emitting gas and the column densities and
  relative abundances of water and other species detected on small
  scales.

\subsection{Water on small scales}\label{sec:smallScales}
The key result from the presented observations is the detection of
 compact \water\ in IRAS2A and IRAS4A-NW, adding to the discovery
towards IRAS4B \citep{jorgensen2010a}. Together, the observations
demonstrate the presence of water emission on 100~AU scales -- much
smaller than probed by the recent \herschel\ \citep{kristensen2010} 
 and \textit{ISO} \citep{nisini1999} observations.

It is particularly noteworthy that the water emission is present in
all these systems --  and the integrated emission (spatially and
  spectrally) is in fact stronger in both IRAS4A and IRAS2A than in
IRAS4B. This is in contrast to the space observations where the
\herschel\ HIFI data show equally strong or stronger emission lines in
IRAS4B than the other sources \citep{kristensen2010} and the
\spitzer\ data only show detections in IRAS4B
  \citep{watson2007}.

The extent of the emission estimated from the sizes of the
  circular Gaussian fits in the $(u,v)$ plane
  (Table~\ref{tab:intensities}) differ between the sources: for IRAS2A
  emission from the different species is generally more extended
  (0\farcs70 and 1\farcs45; 87--181~AU radii) than in IRAS4A-NW
  (0\farcs37 and 0\farcs88; 46--110~AU radii). Still, for both sources
  (as well as IRAS4B), these results show that the bulk of the
  emission is truly coming from the inner 50--100~AU of the
  protostellar systems.

\begin{figure}
    \centering
        \includegraphics[width=\columnwidth]{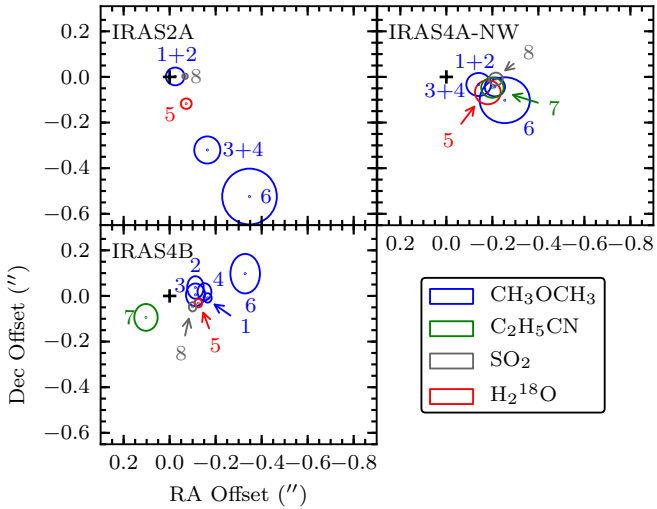}
    \centering
        \caption{\label{fig:linepos} Positions of the lines from
            circular Gaussian fits in the ($uv$)-plane to the
            integrated line emission. The positions are plotted
            relative to the continuum peak position (cross) and the
            error in position for the continuum peak position is
            roughly the extent of the cross while the extent of the
            ellipses shows the error in position for the
            lines. Ellipses with the same colors represent the same
            molecule, while the numbers refer to a certain molecule
            and transition which has the same number in
            \figref{fig:spectra}, Table~\ref{tab:intensities},
            Table~\ref{tab:molecules} and Table~\ref{tab:linefits}.2. 
            Many of the detected lines are
            systematically offset from the continuum peak.}
\end{figure}

The positions of the emission differ slightly: \figref{fig:linepos}
  shows the peak positions of Gaussian fits in the
  \textit{(u,v)}-plane to the integrated intensity of the detected lines
  relative to the continuum position for the two sources, IRAS2A and
  IRAS4A-NW, from this paper as well as IRAS4B. The size of the
  ellipses show the statistical uncertainty in the Gaussian fits, and
  the numbers correspond to those in \figref{fig:spectra},
  Table~\ref{tab:intensities}, Table~\ref{tab:molecules} and Table~\ref{tab:linefits}.2
  identifying each line. The position of the lines in the different
  sources show a systematic offset, in IRAS4A-NW and IRAS4B the lines
  are all (except \ethylcyanide\ in IRAS4B) offset about 0\farcs2 west
  of the continuum peak. When fitting the weak line of 
\dimethylether\ (line 6) in IRAS2A, an extra Gaussian was added, 
centered on the stronger extended emission to minimize its affect 
on the fit of the line.
 The offsets are  smaller than the beam, but
  still significant given the signal-to-noise of the data. That the
  shifts are physical is further strengthened by the observation that
  they are similar for the observed species toward each source.

  \citet{chandler2005} detect similar offsets in the line emission of
  complex organic molecules relative to the continuum source towards
  IRAS~16293-2422 combining high-resolution SMA observations with
  archival VLA data. They argue that an outflow shock eastward of one
  component in the protostellar binary (``source A'') is responsible
  for enhancing the abundances, and sputtering the ices off the
  grains. Because the offsets seen here are not aligned with any
  larger scale outflow, and because a tentative velocity gradient
  perpendicular to the outflow is seen towards IRAS4B
  \citep{jorgensen2010a}, we argue that the lines detected here do not
  have their origin in the outflow. The line widths of
  $1.6-4.1$~\kms\ observed in these data (Sect.~\ref{sec:masing}) are
  also significantly less than the outflow widths seen in
  \herschel-HIFI spectra \citep{kristensen2010}.

\subsection{Water associated with the outflow in IRAS2A}


Whereas the bulk of the H$_2^{18}$O emission toward the three
  sources is associated with the central protostellar component, some
  extended emission at the velocity of the H$_2^{18}$O line is seen
  extending toward the southwest in the integrated intensity map for
  IRAS2A (\figref{fig:mom0}). Previous observations of IRAS2A have
  shown that it is oriented such that the outflow lobe extending in the
  SW direction is blueshifted \citep{jorgensen2004,jorgensen2007}.

\begin{figure}[h]
    \centering
    \includegraphics[width=\columnwidth]{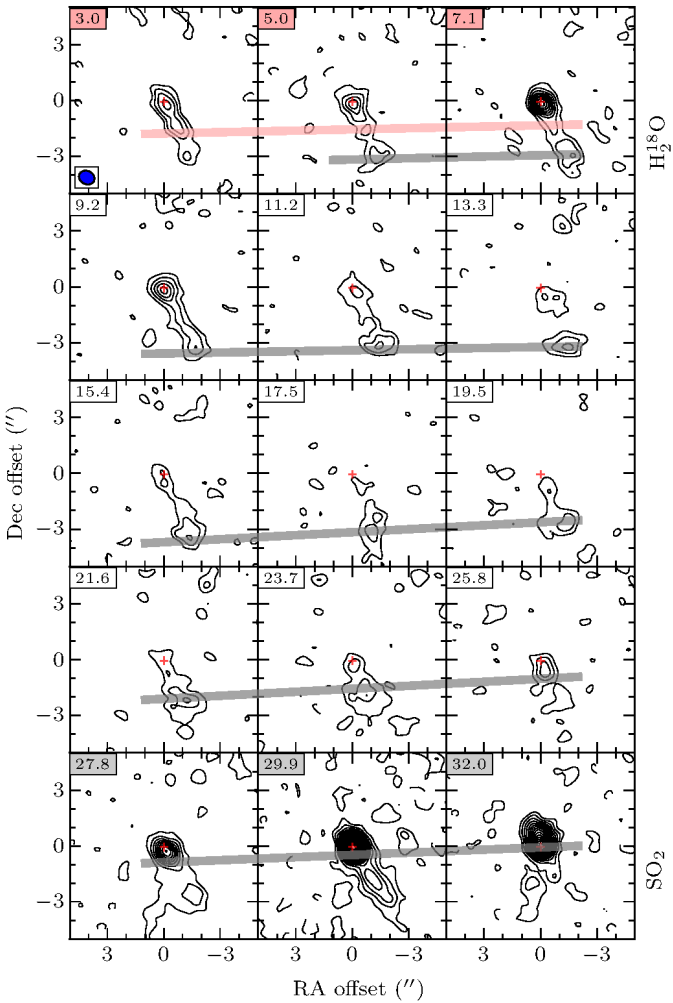}
    \centering
    \caption{\label{fig:i2a_chmap1} Continuum subtracted channel map
      between 2 and 33~\kms\ for IRAS2A, velocity in \kms\ given in
      each frame, channel width is 2.1~\kms. Contours in steps of
      3$\sigma$ (24~mJy~beam$^{-1}$~\kms) from a 3$\sigma$ level. The
      beam size is given in the lower left corner of the first
      frame. Gray and red broad lines trace the different outflow
      components of \sulfurdioxide\ and \water, respectively . }
\end{figure}

To investigate this emission further, continuum subtracted
  channel maps between 2 and 33~\kms, integrated in steps of 2.1~\kms,
  are shown in \figref{fig:i2a_chmap1}. Starting at the higher
  velocities, corresponding to \sulfurdioxide\ emission in
  \figref{fig:spectra}, strong on-source emission is seen, but also
  the extended outflow emission. The gray lines trace the rough
  position of the peak outflow emission in the frames 5.0 to
  32.0~\kms, corresponding to a linear velocity gradient
  characteristic for outflow affected material: the peak of the
  emission in each frame moves further away along the outflow when
  moving to more blue-shifted velocities. At 19.5~\kms\ the knot
  identified as \sulfurdioxide\ has moved about 3\arcsec\ away from
  the central source. The knot persists at this position
  (3\textsim4\arcsec\ away) almost throughout the rest of the maps and
  at a velocity of 5.0--7.1~\kms\ it weakens. At
  9.2~\kms\ emission is seen to emerge close to the protostar, which
  is identified as \water. It strengthens slightly and moves away from
  the source with lower velocities. Just as with the
  \sulfurdioxide\ emission it can be traced, although over a smaller
  number of channels and spatial extent (red line). The outflow water
  emission is extended and weak, but shows that water is present in
  the outflow of IRAS2A.

\begin{figure}
  \centering
    \includegraphics[width=\columnwidth]{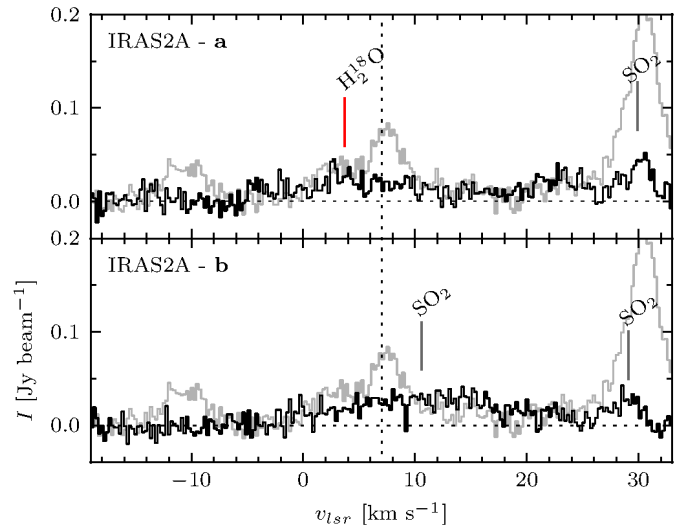}    
    \caption{\label{fig:offpos_spectra} IRAS2A off-position spectra,
      averaged over 1\arcsec$\times$1\arcsec\ regions around the
      positions marked with {\bf a} and {\bf b} in \figref{fig:mom0}
      in the blueshifted outflow from the imaged data. The spectra
      have been rebinned to twice the velocity resolution (width of
      0.23~\kms). The systemic velocity of the \water\ line is drawn
      with a vertical dotted line at $v_{\rm lsr}=7$~\kms, and
        the gray outline in the background is the on-source spectrum
        averaged in a region of the same size as the other spectra.}
\end{figure} 

The detection of extended outflow \water\ and
  \sulfurdioxide\ emission is further strengthened when looking at
  spectra at different positions in the outflow. Spectra extracted at
  two positions (averaged over 1\arcsec$\times$1\arcsec\ regions) from
  the low surface brightness material in the outflow are shown in
  \figref{fig:offpos_spectra}. In \figref{fig:mom0} the positions are
  marked with {\bf a} (magenta cross) and {\bf b} (green
  cross). Outflow position {\bf a} shows one \sulfurdioxide\ component
  with a weak blue tail and a broad \water\ component that is
  blueshifted 4--6~\kms\ from the rest velocity. This blueshift
  coincides with the velocity of the \dimethylether\ lines 3 and 4,
  but the absence of the \dimethylether\ lines 1 and 2, which are
  detected in the on-source spectra, shows that it is indeed
  \water. At position {\bf b} only \sulfurdioxide\ with a broad
  blueshifted component and a narrow component is visible.

Thus the extended outflow emission seen towards IRAS2A in
  \figref{fig:mom0} is at least partly blueshifted \water\ in the
  inner part, close to the source and \sulfurdioxide\ in the outermost
  part of the outflow.

\subsection{Constraints on geometry and dynamics}\label{sec:geometry}
\cite{watson2007} suggested that the difference between IRAS4A
  and IRAS4B could be attributed to either hot water only being
  detectable during a short phase of the life cycle of YSOs or
  alternatively a particularly favorable line of sight in that system.
  Because of the lower dust opacity, protostellar envelopes are
  optically thin at submillimeter wavelengths, which allows detections
  of the water emission lines from the smallest scales towards the
  center of the protostars. The detections towards three of three
  targeted sources \citep[][ and this paper]{jorgensen2010a} therefore
  argue against the first of these suggestions.
 
  The detected velocity gradient in IRAS4B -- which is absent in the
  other sources (\figref{fig:mom0}) -- may be a geometrical effect. A
  simple solution could be that the sources without a velocity
  gradient are observed close to face-on. This would require that IRAS4B with
its detected velocity gradient is observed edge-on, however. This is
not supported by observations of the outflows \citep{jorgensen2007}:
all three sources show well-separated red and blue outflow cones
suggesting at least some inclination of the outflows relative to the
line-of-sight. Also, of the three sources, IRAS4B in fact shows the
most compact outflow: if one assumes similar extents of the outflows
this would suggest that IRAS4B rather than the other sources are seen
more close to face-on.

Another way to deduce the inclination of the outflow with respect to
the plane of the sky is through observations of maser spots in the
outflows of the sources, although they do not seem to agree with
observations of the larger scale outflow in many
cases. \citet{marvel2008} conclude that in IRAS4A-NW, the maser
emission cannot be directly associated with the large scale
outflow. For IRAS4B the masers have been associated with the outflow
\citep{marvel2008,desmurs2009}, but the position angle of the ``maser
outflow'' still differs from that of the CO outflow by some
20--30$^\circ$.

\begin{figure}
    \centering
        \includegraphics[width=\columnwidth]{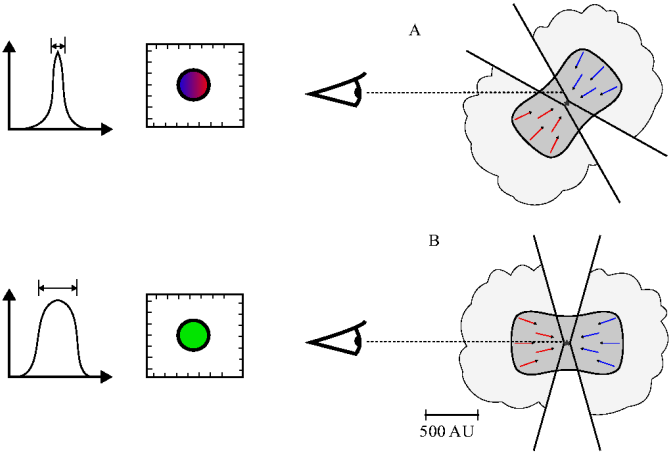}
    \centering
        \caption{\label{fig:drawing} Schematic drawing of a disk
          dominated by inward motions rather than rotation in different
          observation geometries. In {\it case A} (IRAS4B), the observed line
          is narrow and a small velocity gradient is seen. In {\it
            case B} (IRAS2A and IRAS4A-NW), the observed line is broader and no velocity
          gradient is observed.}
\end{figure}

We propose that the compact H$_2^{18}$O emission has its origin in a
flattened disk-like structure where inward motions is still dominating the
dynamics, and where the viewing angle defines the observational
characteristics (\figref{fig:drawing} for a schematic
drawing). In fact, \citet{brinch2009} suggest, based on modeling of submillimeter
observations, that IRAS2A indeed has such a structure on a few 100~AU scales.
Observing a source with this structure at different viewing angles
would potentially produce the observed differences.

In the proposed solution, IRAS4B is viewed at an angle, closer to
face-on, causing the mid-IR emission observed by \spitzer\ to escape
from the envelope through the cavities produced by the outflow, where
the optical depth along the line of sight is lower in agreement with
the suggestion by \citet{watson2007} (case A in
\figref{fig:drawing}). In IRAS2A and IRAS4A-NW, the viewing angle is
closer to edge-on, causing mid-IR emission to be blocked by the
optically thick envelope (case B in \figref{fig:drawing}).


This solution could also explain the velocity gradient observed
towards IRAS4B and the broader line widths towards IRAS2A and
IRAS4A-NW. In disks dominated by inward motions rather than rotation, the
systems viewed at an angle exhibit narrow lines and a velocity
gradient (case A in \figref{fig:drawing}), while systems viewed edge
on show broader lines and no velocity gradient (case B in
\figref{fig:drawing}). The dynamical structure of the disk would move
towards a rotationally supported disk on a local dynamical time-scale,
which is a few~\tento{3}~yr at a radius of 110~AU \citep{brinch2009}.

\subsection{Non-masing origin of H$_2^{18}$O emission}\label{sec:masing}

We now turn the attention to chemical implications of the
  presented data: the subsequent analysis of the data focuses on the
  abundance and excitation of water and other species detected in the
  spectra.

The primary concern about the \water\ analysis in this context is
  whether the \water\ line could be masing. Many lines of water
observable from the ground are seen to be masing, as is the case for
the same transition studied here (\htwootrans) but in the main
isotopologue, H$_2^{16}$O \citep{cernicharo1994}. Several observables
argue against that the \water\ \htwootrans\ line is masing in the
sources investigated here.

First, the line widths observed in the central beam towards the
continuum peak are comparable for all the detected species in each of
the sources, as was also seen in IRAS4B
\citep{jorgensen2010a,jorgensen2010b}. Although the mean line
  width is somewhat broader in IRAS2A and IRAS4A-NW compared to IRAS4B
  they are still significantly more narrow towards the central source
  (\textless5~\kms) than in the broad outflow components seen in the
  \herschel\ data (\textgreater~5~\kms) towards these sources
  \citep{kristensen2010}. For IRAS2A, the mean FWHM and standard
deviation of the fitted lines is \mbox{2.8$\pm$0.7~\kms,} for
{IRAS4A-NW} \mbox{2.4$\pm$1.0~\kms,} and for IRAS4B
\mbox{1.1$\pm$0.2~\kms}. In addition to this, the variations in
intensities between the different lines in each source are smaller
than the variations in line intensity between the sources.

In IRAS2A, \citet{furuya2003} report that maser emission at 22~GHz is
highly variable and has a velocity of 7.2~\kms, i.e. marginally
red-shifted. For IRAS4A, \citet{furuya2003} and \citet{marvel2008}
show that the strongest maser component lies at $\sim$10~\kms\ where
we do not detect any strong \water\ millimeter emission.

Taken together, these arguments suggest that the \water\ emission
detected towards the central positions of the sources is not masing,
but instead is thermal emission with the same origin as the other
detected species.

\subsection{Column densities and abundances determinations}\label{sec:colden}
Having established that the \water\ line is unlikely to be masing, the
column densities for each species can be determined and compared
directly. The emission from the different molecules is assumed to be
optically thin and to uniformly fill the beam (valid since the
deconvolved size in \tabref{tab:intensities} is comparable to the beam
size). The excitation is taken to be characterized by a single
excitation temperature $T_\mathrm{ex}$, which does not have 
to be equal to the kinetic temperature.

\subsubsection{Water column density and mass}\label{sec:water}

\tabref{tab:densities} summarizes the inferred column densities of the
various molecules assuming $T_\mathrm{ex}$=170~K. The gas-phase
\normalwater\ abundance and mass are calculated with an isotopic
abundance ratio of \mbox{\up{16}O/\up{18}O} of 560 in the local
interstellar medium \citep{wilson1994}. The masses of gas phase
\normalwater\ in the sources are given in the bottom of
\tabref{tab:densities}, in units of Solar and Earth masses, along with
the fraction of gas phase water relative to H$_2$. As pointed out by
\citet{jorgensen2010a}, in IRAS4B, the derived column density of
\normalwater\ is almost 25 times higher than deduced by
\citet{watson2007}. The deduced mass of the gas phase water in IRAS2A
is roughly six times higher than in IRAS4A-NW and 90 times higher than
in IRAS4B. Thus, water is present in the inner parts of all of the
sources and the non-detection in IRAS4A and IRAS2A at mid-infrared
wavelengths is due to the high opacity of the envelope combined with a
strong outflow contribution to H$_2^{16}$O lines in IRAS4B \citep[see
  also][]{herczeg2011}.

\begin{table}[h]
    \caption{\label{tab:densities} Derived properties of the detected
      species over the deconvolved sizes of the lines for all sources
      in the project.}
        \begin{center}
\begin{tabular}{lllllll}  
\hline\hline
                             				 	   	& IRAS2A            					& IRAS4A-NW        & IRAS4B                 \\
\hline 
$N_\mathrm{CH_3OCH_3}$\tblmark{a} 	& 1.2\tento{17}\tblmark{b}     		& 6.8\tento{16}    &   5.6\tento{16}   \\
$N_\mathrm{H_2^{18}O}$\tblmark{a}     	& 2.9\tento{16}                         		& 7.9\tento{15}    &   3.9\tento{15}   \\
$N_\mathrm{C_2H_5CN}$\tblmark{a} 		& $<$1.4\tento{14} 				& 1.1\tento{15}    &   2.8\tento{14}   \\
$N_\mathrm{SO_2}$\tblmark{a}               	& 1.8\tento{16}         				& 8.2\tento{15}    &   1.3\tento{15}   \\
\hline
M$_{\rm H_2O}$ (\Msun)                             	&  1.8\tento{-6}        			 	& 2.7\tento{-7}    &  1.5\tento{-8}   \\
M$_{\rm H_2O}$ (\Mearth)                        	&  0.61                          				& 0.09                      &  0.006                  \\
$X$(=$N_{\rm H_2O}/N_{\rm H_2}$)        	&  4.2\tento{-6}         				& 1.4\tento{-7}    & 7.8\tento{-9}   \\
\hline
\end{tabular}
\end{center}
\tablefoottext{a} In cm\up{-2} assuming $T_\mathrm{ex}$=170~K. The
typical error is a combination of the assumption of the excitation
temperature and the flux calibration uncertainty ($\sim$20\%). The
determined column density varies with $T_\mathrm{ex}$ in the interval
100--200~K differently for each molecule, see Fig.~\ref{fig:xvstex}
and the discussion in Sect.~\ref{sec:tex}.
\tablefoottext{b} Using only lines 1, 2, 3 and 4.
\end{table}

\begin{figure}
\centering
    \includegraphics[width=\columnwidth]{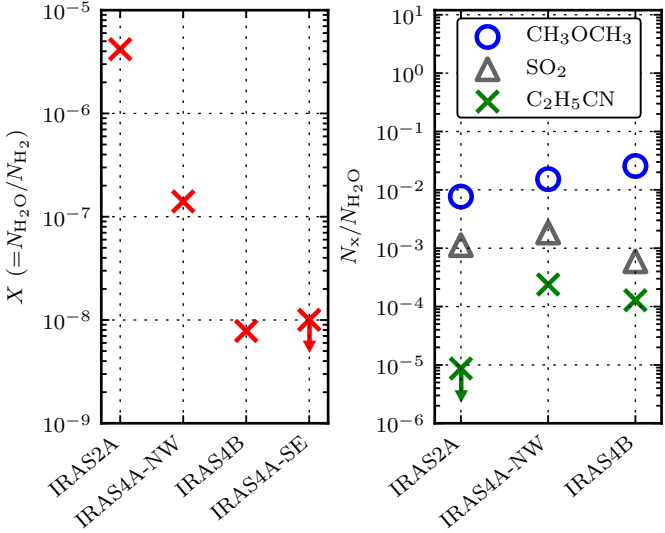}
    \caption{\label{fig:rel_abundances} \textit{Left:} Fraction
        of gas phase water relative to H$_2$ \citep[from the continuum
          observations of][]{jorgensen2009} for the different sources,
        including an upper limit estimate for IRAS4A-SE.  \textit
      {Right:} Ratio of the column density of the molecules relative
      to water. The \ethylcyanide\ value for IRAS2A is an upper
      limit.}
\end{figure}

The ratio of \normalwater\ gas phase abundance to the total H$_2$
abundance (derived from continuum observations \citep{jorgensen2009})
in \figref{fig:rel_abundances} shows that the fractional abundance of
gas phase water in IRAS2A is much higher than in IRAS4A-NW and
IRAS4B. Still, the gas-phase water likely accounts for only a small
fraction of the total amount of water (gas+ice) on those scales.
Typical \normalwater\ to H$_2$ abundance ratios of 10\up{-5} --
10\up{-4} have been deduced in shocked regions close to
protostars \citep{harwit1998,nisini1999}, and in ices in the cold
outer parts of protostellar envelopes \citep{pontoppidan2004}. In
shocks, high temperature gas-phase reactions drive much of the oxygen
into water \citep{draine1983}. In more quiescent regions the ice is 
expected to evaporate off the grains
when heated to temperatures above 100~K in close proximity to the
central source \citep{fraser2001}. The fact that our inferred
abundance ratios are generally much lower than 10\up{-4} suggests that
a large fraction of the dust mass in the inner $\sim$100~AU has a
temperature below 100~K, with the fraction varying from source to
source. In colder regions, most of the water gas originates from
UV photodesorption of ices. Assuming a total \normalwater\ to H$_2$ abundance 
ratio of
10\up{-4} (gas+ice), the fraction of water in the gas phase is 4\% for IRAS2A,
0.1\% for IRAS4A-NW and 0.01\% for IRAS4B of the total water content
(gas+ice). This difference could be due, for example, to varying
intensities of the UV radiation from the newly formed stars and
geometry causing differences in the relative amounts of material
heated above the temperature where water ice mantles are evaporated.
Towards the more evolved Class II source TW Hydra,  \citet{hogerheijde2011} deduce
a gas phase water to total hydrogen abundance of 2.4\tento{-8} using \herschel. 
This value is of the same order of magnitude as our results, and places TW Hydra just 
above IRAS4B in \figref{fig:rel_abundances}.

\subsubsection{Other molecules}\label{sec:other}

In the right plot of \figref{fig:rel_abundances}, the abundances of
the detected molecules relative to \normalwater\ are presented. The
\dimethylether\ and \sulfurdioxide\ content relative to \water\ is
roughly comparable between the different sources.  This suggests 
that for \dimethylether, \sulfurdioxide\ and \water\ (and thus \normalwater) 
similar chemical processes are at work in the three sources.
 \ethylcyanide\ is the least abundant of the molecules
for all the sources. In IRAS2A the high abundance of water is
contrasted by the relatively low abundance of \ethylcyanide, for which
only an upper limit is deduced. This relatively low abundance in
IRAS2A is robust even when much lower excitation temperatures are
assumed (see Sect.~\ref{sec:tex}).

It is also evident from \figref{fig:rel_abundances} and
\tabref{tab:densities} that the \dimethylether/\ethylcyanide\ column
density ratio varies significantly between the sources. This ratio is
10\up{3} for IRAS2A, 60 for IRAS4A-NW and 200 for IRAS4B assuming
$T_\mathrm{ex}$=170~K. The differences in relative abundances of N and
O-bearing complex organics have been interpreted as an indicator for
different physical parameters \citep{rodgers2001} and different
initial grain surface composition
\citep{charnley1992,bottinelli2004}. \citet{caselli1993} argue that
the (dust) temperature history of the cloud, in particular whether it
was cold enough to freeze out CO, is an important factor for the
resulting ratio.

 \subsubsection{Effect of excitation temperature}\label{sec:tex}

The results presented in Sections~\ref{sec:water} and \ref{sec:other}
assume an excitation temperature of 170~K for all species. In this
section we demonstrate that all our conclusions are robust over a wide
range of assumed temperatures. The adopted $T_\mathrm{ex}$=170~K 
is the same as found from the model
fit to the mid-infrared water lines observed by \spitzer\
\citep{watson2007} and allows for easy comparison to this paper and
\citet{jorgensen2010a}. 

Although the origin of the mid-infrared and millimeter lines are
likely not the same, this temperature is a reasonable first guess
appropriate of the conditions in the inner 150~AU of protostellar
systems. From modeling of dust continuum observations, the density in
the inner envelope/disk is found to be high, at
least 10\up{7}~cm\up{-3}, and the dust temperature is thought to be at
least 90~K for a fraction of the material
\citep[e.g.,][]{jorgensen2002,jorgensen2004}. This is also the
temperature at which water is liberated from the grains; any gas-phase
water detected in the high excitation line observed here
($E_\mathrm{u}$$\approx$200~K) likely originates from the warmer
regions (Sect.~\ref{sec:water}).

The current observations and literature data do not provide enough
information on the line emission at these scales to constrain the
excitation temperature directly. For none of the detected species are
there more than two previous measurements and those detections either
have large uncertainties due to their weakness \citep{jorgensen2005}
or are sensitive to larger scale emission including outflow and
envelope because of the large beam
\citep{bottinelli2004,bottinelli2007,wakelam2005}. For the deeply
embedded low-mass protostar IRAS~16293-2422, \citet{bisschop2008}
derive rotational temperatures between 200--400 K from several compact
emission lines of complex molecules using SMA data. Furthermore, for
the outflow toward IRAS4B \citet{herczeg2011} deduce
$T_\mathrm{ex}$=110--220~K with recent \herschel\ PACS data.
 
\begin{figure}
\centering
    \includegraphics[width=\columnwidth]{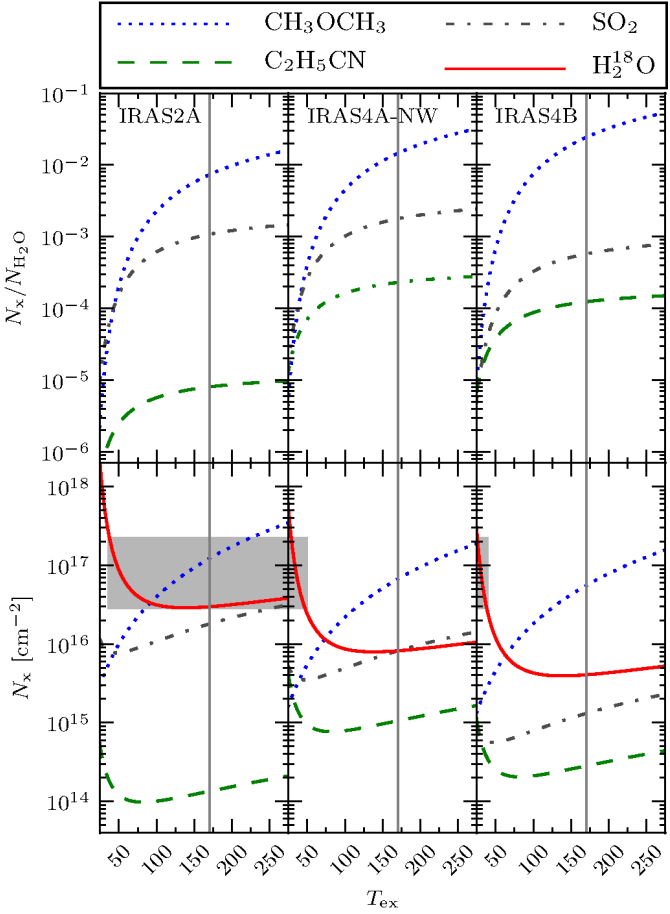}
    \caption{\label{fig:xvstex} Plots showing (\emph{upper}) the
        ratio between the detected molecules and water (\normalwater), 
        assuming that the species have the same $T_\mathrm{ex}$
        and (\emph{lower}) the column density of all the detected
        lines (including \water). The colors and line style indicate
        the different molecules, the gray vertical line mark
        $T_\mathrm{ ex}$=~170~K and the gray area in each plot
        indicate the temperatures where the column density of
        \water\ is the same for the sources.}
\end{figure}

Figure~\ref{fig:xvstex} explores what happens with the column
densities at significantly different excitation temperatures than
those assumed here. It shows two different plots for the three
sources. The plots in the top row show the ratio between the column
density of the various molecules and that of water, whereas the plots
in the bottom row show the column densities themselves, all as
functions of excitation temperature. The ratios
($N_\mathrm{X}/N_\mathrm{H_2O}$) do not change significantly between
100 and 275~K. The gray area in the lower plots shows the range of
excitation temperatures for which the column density of water is the
same in the three sources. Only if the excitation temperature in
IRAS4A-NW and IRAS4B is lower than 50~K can all sources have comparable
water column densities based on the observed intensities.

\begin{figure}[h]
\centering 
    \includegraphics[width=\columnwidth]{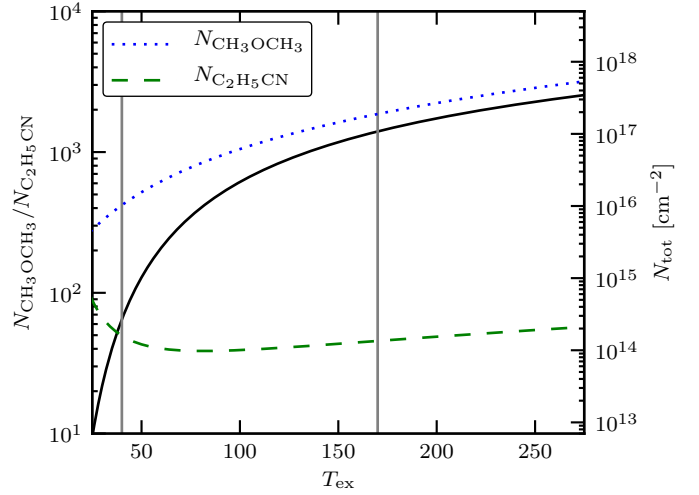}
    \caption{\label{fig:ratio_i2a} Plot of the ratio between
      \dimethylether\ and \ethylcyanide\ in IRAS2A (left $y$-axis,
      filled line), and the column densities derived for those
      molecules (right $y$-axis, dashed/dash dotted line) for
      temperatures between 25 and 275~K. The vertical lines show
      {$T$=40 and 170~K}.}
\end{figure}

The different molecules could of course each have different excitation
temperatures. Specifically, can different excitation temperatures be
the cause of the large range in inferred
\dimethylether/\ethylcyanide\ column density ratios?
Figure~\ref{fig:ratio_i2a} shows this ratio and the individual column
densities for excitation temperatures from 25 to 275~K for IRAS2A. The
corresponding plots for the other two sources show similar trends (see
also \figref{fig:xvstex}). The temperature does not affect the column
density estimate for \ethylcyanide\ significantly, showing that our
upper limit estimate is rather robust, and that the differences in
this ratio between the sources cannot be attributed to differences in
the excitation of \ethylcyanide\ alone.  On the other hand, changing
the excitation temperature for \dimethylether\ affects the column
density and the ratio to a greater extent. This difference in behavior
with $T_\mathrm{ex}$ for the two molecules is caused by the
transitions' upper energy levels ($E_\mathrm{u}$); 18~K for
\dimethylether\ versus 122~K for \ethylcyanide. For \dimethylether,
its lower value of $E_\mathrm{u}$ causes the column density estimate
to decline more rapidly with $T_\mathrm{ex}$ than for \ethylcyanide:
the contribution from the exponent in the Boltzmann distribution
(exp(--$E_\mathrm{u}$/k$T_\mathrm{ex}$)) does not compensate for the
fall that the partition function (proportional to
$T_\mathrm{ex}$\up{b}, where $b$\textgreater0) introduces. In the
temperature range 40 to 250~K the \dimethylether/\ethylcyanide\ column
density ratio varies with a factor of about 30. Looking only at the
\dimethylether/\ethylcyanide\ column density ratio, the excitation
temperature for IRAS2A has to be $T_\mathrm{ex}$=45~K, and for IRAS4B
$T_\mathrm{ex}$=85~K, to match that of IRAS4A-NW.

As argued above based on the physical characteristics of the sources,
values of $T_\mathrm{ex}$ significantly below 100~K and large
differences in $T_\mathrm{ex}$ between molecules and sources are not
expected on the scales probed here.  Also, the agreement in widths,
extent and positions between the lines argues against this.  If the
temperature is allowed to vary in the range 100-250 K the conclusions
drawn in this paper are unchanged.

\subsection{Formation of molecules in the protostellar environment}\label{sec:formation}

The simultaneous observations of water and the complex mole\-cules can
also be used to shed new light onto where and how the formation of
various molecules takes place in the protostellar environment. The
basic scenario is that many of the complex organic molecules are
formed as a result of grain-surface reactions during the collapse of
the protostellar envelopes as material is heated slowly
\citep[see][for a review]{herbst2009}.

 
\citet{garrod2006} use a gas-grain network of reactions to investigate
the formation of certain complex organic molecules in hot cores and
corinos during the warm-up and evaporative stages. The results show
that the chemistry occurs both on grain surfaces and in the gas phase,
with the two phases being strongly coupled with each other. Large
abundances of complex organic molecules are present at the end of this
warm-up stage, and show that a period of 10\up{4} -- 10\up{5} years is
vital for the formation of complex molecules. 

A key question is thus whether the material in these low-mass
protostars spends enough time at moderate temperatures of 20--40~K to
form complex organic molecules during the collapse of the cloud and
formation of the disk. A parcel of material originating in the cold
core will pass through the warm inner envelope during the collapse and
be incorporated into the disk before being accreted onto the
star. Depending on its precise origin, the material can follow many
different routes, but exploration of a variety of collapse models
shows that most parcels indeed spend the required period at moderate
temperatures to form the complex organics, either in the inner
envelope or in the disk \citep{visser2009}. Some of the routes result
in the material being directly exposed to UV radiation near the
outflow cavity walls \citep{visser2011}.  After
evaporation, gas-phase temperatures in the hot core are generally too
low \citep[i.e., 
$T\lesssim230$~K,~$R\gtrsim20$~AU,][]{larson1972,jorgensen2002} for
any significant gas-phase formation of water to take place. The fact that 
we observe narrow emission lines of complex molecules
on small scales confirms that some fraction of the collapsing
cloud has spent enough time under conditions that lead to the formation of complex
species that will enter the disk.









\section{Summary and Outlook}\label{sec:sum}

We have presented subarcsecond resolution observations of the
\water\ \htwootrans\ transition at 203.4~GHz towards a small selection
of deeply embedded protostars in the NGC~1333 region. In
  addition to the targeted \water\ molecule, \sulfurdioxide\ and
  complex organic molecules are seen.

\begin{itemize}

\item Water is successfully detected towards IRAS2A and one component
  of the IRAS4A binary. These data add to the previous detection of
  the \htwootrans\ transition of \water\ towards IRAS4B by
  \citet{jorgensen2010a}. The origin of the emission differs
  from that seen with both \herschel\ and \spitzer, where 
 larger-scale emission (protostellar envelope, outflow shocks/cavity walls)
  is detected.  

\item The gas-phase water abundance is significantly lower than a
  canonical abundance of 10\up{-4}, suggesting that most water is
  frozen out on dust grains at these scales. The fraction of water in
  the gas phase is 4\% for IRAS2A, 0.1\% for IRAS4A-NW and 0.01\% for
  IRAS4B of the total water content (gas+ice).

\item IRAS2A shows \water\ eission not only at the source position,
  but also in the blueshifted outflow. In IRAS4A we only detect water
  towards IRAS4A-NW. None of the lines presented for the other sources
  are detected towards IRAS4A-SE.

\item Based on the small species-to-species variations, in both line
  strength and line width, we argue that the
  \water\ \htwootrans\ emission line is not masing in these
  environments and on these scales.

\item While IRAS4B shows narrow emission lines and a tentative
  velocity gradient, reported by \citet{jorgensen2010a}, the other
  sources show broader lines and no velocity gradients. Additionally,
  IRAS4B is the only source of the three detected in mid-infrared
  water lines by \citet{watson2007}. We propose that a disk-like structure
  dominated by inward motions rather than rotation with different orientations
  with respect to the observer could explain the observed line
  characteristics.

\item The abundances of the detected molecules differ between the
  sources: the abundance of \ethylcyanide\ with respect to water,
  compared to both \sulfurdioxide\ and \dimethylether, is much lower
  in IRAS2A than the other sources. This shows that the
    formation mechanism for \ethylcyanide\ is different from the other
    molecules, indicating that it is not trapped in the water ice
    mixture or related to the sublimation of water.

\end{itemize}

The detections of \water\ emission from the ground show good promise
for using \water\ as a probe of the inner regions of protostars and
for studying the formation of complex organic molecules in the
water-rich grain ice mantles. Higher resolution observations, using
the PdBI in the most extended configuration, or with the Atacama Large
Millimeter/submillimeter Array (ALMA) band 5 could be used to reveal
the kinematical structure of the 10--100~AU scales and the dynamics of
the forming disks. Observations of higher excited lines of H$_2^{18}$O 
and the complex organics in other ALMA bands (e.g., band~8) 
could provide a better handle on the excitation conditions of water 
on few hundred AU scales in the centers of protostellar envelopes. 
Finally, observing more protostellar sources in the 
\water\ \htwootrans\ transition would provide a better statistical 
sample for comparisons and conclusions.

\begin{acknowledgements} We wish to thank the IRAM staff, in
  particular Arancha Castro-Carrizo, for her help with the
  observations and reduction of the data. The research at Centre for
  Star and Planet Formation is supported by the Danish National
  Research Foundation and the University of Copenhagen’s programme of
  excellence. This research was furthermore supported by a Junior
  Group Leader Fellowship from the Lundbeck Foundation to JKJ and by a
  grant from Instrumentcenter for Danish Astrophysics. This work has
  benefited from research funding from the European Community's sixth
  Framework Programme under RadioNet R113CT 2003
  5058187. Astrochemistry in Leiden is supported by a Spinoza grant
  from the Netherlands Organization for Scientific Research (NWO) and
  NOVA, by NWO grant 614.001.008 and by EU FP7 grant 238258.
\end{acknowledgements}

\bibliographystyle{aa} \bibliography{bibfile}


\begin{appendix} 

\section{Tables}
\begin{table}[h]
\label{line:data}
\begin{minipage}{0.79\linewidth}
    \caption{\label{tab:molecules} Tabulated data for the different
      lines detected from JPL \citep{pickett1998} and CDMS
      \citep{muller2001}. }

\begin{tabular}{lllllllll}
\hline\hline                                                                             
No.      & Molecule        & Transition                     & Rest Freq. & $E_\mathrm{u}$ & $\log{A_{ij}}$      \\
         &                 &                                & (GHz)      & (K)       &   ($\log s^{-1}$)  \\
\hline                                                                                                 
1        &\dimethylether  & 3\down{3,0} -- 2\down{2,1} EE  & 203.4203   & 18.1       &  -4.61            \\
2        &\dimethylether  & 3\down{3,0} -- 2\down{2,1} AA  & 203.4187   & 18.1       &  -4.23            \\
3        &\dimethylether  & 3\down{3,0} -- 2\down{2,1} AE  & 203.4114   & 18.1       &  -4.23            \\
4        &\dimethylether  & 3\down{3,1} -- 2\down{2,1} EE  & 203.4101   & 18.1       &  -4.46            \\
5        &\water          & 3\down{1,3} -- 2\down{2,0}     & 203.4075   & 203.7        &  -5.32            \\
6        &\dimethylether  & 3\down{3,1} -- 2\down{2,1} EA  & 203.4028   & 18.1       &  -4.39            \\
7        &\ethylcyanide   & 23\down{2,22} -- 22\down{2,21} & 203.3966   & 122.3        &  -3.15            \\
8        &\sulfurdioxide  & 12\down{0,12} -- 11\down{1,11} & 203.3916   & 70.1       &  -4.07            \\
\hline
\end{tabular}\\
\end{minipage}
\end{table}

\begin{table*}[h]\label{tab:linefits}
\centering

    \caption{Position and FWHM of the detected lines from Gaussian fits to the spectra.}

\begin{tabular}{llccccccc}
\hline\hline
	&	 			&				&	IRAS2A           	&			& &			  	&  IRAS4A-NW	& 		\\
No. 	&	Molecule 		&	Position		&	Width		&  Interval\tblmark{a}	& &  Position	       	&	Width	& Interval\tblmark{a}		\\
 	&		 		&	(\kms)	        	&	(\kms)		&	(\kms)	& &  (\kms)	        	&	(\kms)   	& (\kms)		\\
\hline
1   	& \dimethylether	&	$-11.7\pm0.1$	&	$2.3\pm0.1$	& $-14.0--7.0$ 	& & $-11.3\pm0.1$ 	&$1.6\pm0.1$	& $-13.0- -7.3$	\\
2   	& \dimethylether	&	$-9.3\pm0.1$	&	$2.3\pm0.1$	&	$-$		& & $-9.0\pm0.1$ 	& $1.6\pm0.1$	& $-$	 \\
3   	& \dimethylether 	&	$+1.3\pm0.1$	&	$2.3\pm0.1$	& $-1.0-+4.4$ 	& &  $+1.7\pm0.1$	& $1.6\pm0.1$	&  $0.0-+4.7$	\\
4   	& \dimethylether	&	$+3.2\pm0.1$	&	$2.3\pm0.1$	&	$-$		& &  $+3.6\pm0.1$	&$1.6\pm0.1$  	& $-$	\\
5   	& \water 	 		&	$+7.4\pm0.1$	&	$4.0\pm0.1$	& $+4.4-+11.4$ 	& &  $+6.6\pm0.1$	&$2.9\pm0.3$   	& $+4.7-+9.6$	\\
6   	& \dimethylether	&	$+15.2\pm0.2$	&	$2.3\pm0.1$	& $+12.9-+17.5$ & &  $+14.4\pm0.1$	&$1.6\pm0.1$  	& $+12.0-+15.2$	\\
7   	& \ethylcyanide	 &	$-$			&	$-$			& $-$ 		& &  $+22.8\pm0.2$	&$3.8\pm0.4$  	& $+19.0-+26.9$	\\
8   	& \sulfurdioxide	 &	$+30.4\pm0.1$	&	$3.8\pm0.1$	& $+26.6-+33.9$	& &  $+30.4\pm0.1$	&$4.1\pm0.2$  	& $+26.9-+34.6$	\\
\hline
\end{tabular}\\
\tablefoottext{a}{Interval over which the integrated line intensity (Table~\ref{tab:intensities}) is estimated.}

\end{table*}

\end{appendix}

\end{document}